\documentclass[12pt]{iopart}
\usepackage{iopams}
\usepackage[dvipsnames]{xcolor}
\usepackage{amsmath}
\usepackage{amssymb}
\usepackage[normalem]{ulem}
\usepackage{cite}

\usepackage{booktabs} 

\usepackage{hyperref}
\hypersetup{
  colorlinks   = true,    
  urlcolor     = blue,    
  linkcolor    = blue,    
  citecolor    = blue      
}
\usepackage{orcidlink}

 \usepackage{etoolbox}
 \usepackage{tikz}

\usepackage{graphicx}
\usepackage[utf8]{inputenc}
\usepackage{appendix}

\begin{document}
\title{Assessing the potential of LIGO-India in resolving the Hubble Tension}

\author{Kanchan Soni$^{1,}$$^2$\orcidlink{0000-0001-8051-7883}, Aditya Vijaykumar$^{3,}$$^4$\orcidlink{0000-0002-4103-0666}, Sanjit Mitra$^{2}$\orcidlink{0000-0002-0800-4626}}

\address{$^1$ Department of Physics, Syracuse University, Syracuse, NY 13244, USA}
\address{$^2$ Inter-University Centre for Astronomy and Astrophysics (IUCAA), Post Bag 4, Ganeshkhind, Pune 411 007, India}
\address{$^3$Canadian Institute for Theoretical Astrophysics, University of Toronto, 60 St George St, Toronto, ON M5S 3H8, Canada}
\address{$^4$ International Centre for Theoretical Sciences, Tata Institute of Fundamental Research, Bangalore 560089, India}

\ead{ksoni01@syr.edu}

\begin{abstract}

Conclusive determination of the Hubble constant ($H_0$) is a major challenge in current astronomy due to the observed discrepancies between early and late universe measurements. Joint detections of gravitational waves (GW) and electromagnetic (EM) signals from binary neutron star (BNS) mergers promise an independent probe for $H_0$. However, this requires tens of such detections, requiring decades of observation with the present detectors. LIGO-India can significantly accelerate this process. LIGO-India's addition to the detector network could boost event detections by $\sim 70\%$ and double the number of properly localized (``triple-coincidence'') detections. Through end-to-end simulations, we show that LIGO-India could increase the EM follow-up rate of kilonovae by $\sim 2-7$ times, reducing the $H_0$ estimation error by a factor of $\approx 1.65-2.82$ for Vera Rubin LSST, thus significantly reducing the observation time required to attain the necessary precision. Moreover, LIGO-India can improve sky localization precision by many folds ($\sim 5$), allowing much deeper EM follow-ups, potentially reducing the time needed to resolve the ``Hubble tension'' from decades to a few years.

\end{abstract}

%
%
%
%
%

\section{Introduction}

The quest for an accurate and precise determination of the Hubble constant ($H_0$), a fundamental parameter that governs the rate of cosmic expansion, has long been a central challenge within cosmology. Despite numerous advances in observational technology, the measurement of $H_0$ remains a subject of ongoing debate. While measurements from the early universe, such as those obtained from measurements of the cosmic microwave background~\cite{Ade_2016,Planck:2018vyg}, consistently yield a lower $H_0$ value with a precision of $< 1\%$, late universe observations, relying on supernovae~\cite{riess2016,Riess_2019,Riess:2021jrx, Brout:2022vxf}, consistently produce a higher value with a precision of $\sim2\%$. If not attributable to systematic errors or statistical fluctuations, this persistent tension raises the possibility of new physics beyond the cosmological standard model. In this context, novel estimates derived from gravitational waves (GWs) emitted by binary systems hold promise to resolve this longstanding \textit{Hubble tension} and refine our understanding of the universe's expansion dynamics.  

Merging binaries are known to be standard sirens for measuring $H_0$. Schutz's original idea~\cite{schutz:1986gp} of estimating $H_0$ by GWs from coalescing binaries led to the foundation of this elusive goal. The direct detection of GWs from the binary merger reveals its apparent luminosity. This luminosity, which is a function of the chirp mass and frequency, can determine the luminosity distance to the source through measurements of the GW source parameters. However, retrieving the redshift information is tedious and cannot be uniquely obtained from the GW signal itself on a single event level. For a coalescing neutron star (NS) with the detection of an electromagnetic (EM) counterpart from the source, e.g., a kilonova, measuring the redshift of the host galaxy and hence $H_0$ becomes more straightforward. The precision of $H_0$ measurement through this method enhances as estimates from more and more such events are stacked. In the absence of counterparts, like in the case of binary black hole mergers or \textit{dark sirens}, statistical association of the redshift can be performed by comparing the GW localization volume with galaxy catalogs~\cite{PhysRevD.77.043512,PhysRevD.86.043011}, or cross-correlating spatial clustering of the GW sources with galaxies~\cite{PhysRevD.93.083511,Bera:2020jhx,Mukherjee:2020hyn ,Mukherjee:2022afz}. Other indirect statistical methods include \textit{spectral sirens}~\cite{Chernoff_1993,PhysRevD.85.023535,Farr:2019twy}, where the redshift of the source is inferred using features in the astrophysical mass distribution of GW sources, and \textit{love sirens}, which breaks the mass-redshift degeneracy via information in the tidal parameter in systems containing neutron stars~\cite{Messenger_2012}.

To date, GW170817~\cite{gw170817_event} is the only binary neutron star (BNS) merger whose EM counterpart has been convincingly observed, marking the beginning of the multimessenger GW astronomy era. Following a short gamma-ray burst known as GRB 170817A~\cite{Abbott_2017,Goldstein_2017,Savchenko_2017} and subsequent observations in various wavebands to detect associated counterparts~\cite{Margutti_2017}, the event's localization with an accuracy of tens of square degrees enabled the determination of its host galaxy, NGC 4993. The measured redshift of this galaxy, combined with the luminosity distance obtained from the GW observation, was then used to constrain $H_0$~\cite{Chen_ish_cosmo_2024}. GW estimates from the Advanced LIGO~\cite{asi} and Advanced Virgo~\cite{advancedvirgo} detectors combined with EM follow-up observations of GW170817 yielded an independent measurement of the $H_0$ value as $70^{+12}_{-8}$ Gpc$^{-3}$yr$^{-1}$~\cite{hubble_constraint_ligo}. Another BNS merger GW190425~\cite{GW190425} was observed in the Advanced LIGO's and Advanced Virgo's third run. Unfortunately, no coincident EM counterpart was observed for this event, partly due to the poor localization of the event, as only two detectors, LIGO Livingston and Virgo, were operating at that moment.

To accurately localize and identify the host galaxies, it is imperative to conduct follow-up observations of the triggering GW events using EM telescopes. However, the feasibility of such follow-up observations is often hindered by physical constraints, such as limited telescope time availability, making it difficult to observe the fast-fading events within a restricted timeframe. This phenomenon may occur if the sources are oriented face-on with respect to Earth or are located at minimal distances, as observed in the case of GW170817. Nevertheless, the probability of such events happening is very low on the basis of our present understanding of astronomy and the last few years of GW observations. Also, the effective duty cycle of the network~\cite{Schutz:2011tw}, the fraction of time the network produces useful data for such observations, which is essentially proportional to the detection rate, remains low. To increase the detection rate for such events, (i) the horizon distance (that is, the sensitivity of the network) should increase, (ii) sky localization should improve so that deeper EM follow-up can be performed, (iii) more detectors in the network, which will increase the effective duty cycle, and (iv) more Target-of-Opportunity time and suitable EM observatories for follow-ups.

In this context, the anticipated participation of LIGO-India in future observation runs, along with the two LIGO detectors at Hanford and Livingston and the Virgo detector, will substantially increase the prospects of detecting more BNS mergers along with their EM counterparts. Several works have previously demonstrated how LIGO-India will improve the sky localization of BNS mergers~\cite{dcc_public_scitific_benefits_of_LI,ligoindiaFairhurst,Saleem_2022_ligoindia,Nissanke_2013}. A recent study~\cite{PhysRevD.109.044051} showed that the inclusion of LIGO-India, with an optimistic duty cycle of 80\%, in a four-detector configuration, can reduce the median sky localization area to approximately 2.4 square degrees. Moreover, future EM observatories, such as the Vera C. Rubin Observatory Legacy Survey of Space and Time (LSST)~\cite{Goldstein_2015,froster_2021}, will observe kilonovae from BNS mergers scanning the sky over a few square degrees in a given observation time~\cite{winter_2020}. Such rapid follow-up of kilonovae would increase the probability of identification of host galaxies, facilitating precise measurement of $H_0$.

Ref.~\cite{chenmaya} demonstrated that achieving a 2\% precision in measuring $H_0$ would necessitate detecting a few tens of BNS mergers along with their EM counterparts. This should be sufficient to resolve the Hubble tension. At the current detection rate, without LIGO-India, even if we observed one such event per year, it would take several decades to achieve the necessary precision. In this article, we assess the importance of LIGO-India in reducing the number of years needed to achieve adequate precision in measuring the Hubble constant. While the effectiveness of the present GW detector network~\cite{Kiendrebeogo:2023hzf} and LIGO-India~\cite{Saleem_2022_ligoindia} in astronomy has been discussed, here we specifically address the potential role of LIGO-India in resolving the Hubble tension, with available details and as much quantitative analysis as possible. To address this point, we perform an end-to-end simulation mimicking realistic scenarios of future observations for an astrophysically motivated population of BNS with LIGO-India and EM follow-ups. We investigate detection rates across varying duty cycles for a four-detector network, using \texttt{PyCBC Live} for signal recovery and the \texttt{gwemopt} toolkit for EM observations via telescopes such as the Zwicky Transient Facility (ZTF)~\cite{Bellm_2019,ztf_Dekany_2020}, the Wide-field Infrared Transient Explorer (WINTER)~\cite{winter_2020}, and LSST. By simulating kilonova light curves, we assess their detectability in the EM follow-ups.

Our paper is structured as follows. We start with back-of-the-envelope estimates to assess the detection probabilities of BNS mergers and improvements in sky localization when LIGO-India is included in a three-detector network in Sec.~\ref{sec:boe}. In Sec.~\ref{sec:method}, we detail the data generation process for simulating a network comprising four detectors—LIGO Hanford, LIGO Livingston, Virgo, and LIGO-India, all operating at A+ sensitivities~\cite{aplus_,VIRGO:2023elp}. Additionally, we discuss injecting realistic populations of BNSs into the data and their recovery using the low-latency search pipeline \texttt{PyCBC Live}. Section~\ref{sec:pycbclive} outlines the selection criteria for detecting a GW event in the simulated data under different duty cycles. In Sec.~\ref{sec:simulating_lightcurve}, we elaborate on generating kilonova lightcurve models for localization purposes. Section~\ref{sec:em_followup} describes the methodology for following up GW events detected in Sec.~\ref{sec:pycbclive}. The findings of our work are presented in Sec.~\ref{sec:results}, and we conclude our paper in Sec.~\ref{sec:conclusion}.

\section{Back-of-the-envelope estimates}\label{sec:boe}

We first motivate the expected improvements when LIGO-India is added to the network of two LIGO and the Virgo detector by performing simple estimates\footnote{Since the KAGRA detector~\cite{akutsu2021overviewkagracalibrationdetector} currently has significantly less sensitivity than the LIGO and Virgo detectors, we do not include it in the estimations here. However, the same estimation procedure can be followed with the inclusion of KAGRA. We mention the effect of including KAGRA in the estimation at certain places to support the above points.}. The network formed by the LIGO Hanford (H), LIGO Livingston (L), and Virgo (V) detectors is referred to as the HLV network. With the addition of the LIGO-India detector at Aundha, India-- interchangeably referred to as LIGO-India or LIGO-Aundha (A)~\cite{ligo_aundha_coordinates}-- the four-detector network is referred to as HLVA. The network of the three LIGO detectors in Hanford, Livingston, and Aundha is called HLA.

\subsection{Increase in horizon distance} 

We assume that both LIGO detectors have the same sensitivity and that LIGO-India will match their sensitivity at the time of operation. Let us say Virgo reaches $\alpha$ times the ``BNS range''~\cite{obsplan} of LIGO. This means that if for an optimally oriented BNS merger with a certain set of astrophysical parameters, the expected observed signal-to-noise ratio (SNR) in the LIGO detectors is $\rho$, the expected observed SNR in the Virgo detector for a similar event will be $\alpha\rho$. This statement is not rigorous, as $\alpha$ would depend on the astrophysical parameters since the detectors' noise power spectral densities (PSDs) can be significantly different. However, since this work focuses on a narrow range of binary masses that are weakly spinning, considering $\alpha$ as a constant should lead to reasonably good estimates. Then, on average, the network SNR, which is the quadrature sum of the SNRs in individual detectors, would increase for the same astrophysical parameters from $\sqrt{2 + \alpha^2} \rho$ to $\sqrt{3 + \alpha^2} \rho$ with the introduction of LIGO-India to the HLV network. Thus, the average SNR increases by a factor of $\sqrt{(3 + \alpha^2)/(2 + \alpha^2)}$. If detection can be claimed for events when the network SNR is greater than a certain threshold (which, in practice, remains in a very tight range irrespective of the BNS parameters), the observable volume coverage by the network increases by a factor of $[(3 + \alpha^2)/(2 + \alpha^2)]^{3/2}$. If $\alpha=0$, that is, Virgo is not operational, this number becomes $(3/2)^{3/2} \approx 1.84$. If $\alpha=1$, that is, Virgo's sensitivity is comparable to LIGO, the number is $(4/3)^{3/2} \approx 1.54$. If Virgo reaches, say, 70\% (50\%) of the sensitivity of LIGO, the factor becomes approximately $1.66$ ($1.74$). Considering that the projected horizon distance for Virgo corresponds to $\alpha \lesssim 0.7$~\cite{Kiendrebeogo:2023hzf}, {\em LIGO-India is likely to increase the HLV network detection rate by $\sim 70\%$} when all detectors are simultaneously observing. Inclusion of the KAGRA detector, even with the most optimistic projection~\cite{dcc_public_ligo_sensitivity}, would change this number by a few percent (the increase due to LIGO-India in that case would be $\sim 65\%$ or more). Since BNS mergers observation with the present generation detectors is limited to a maximum of a few hundred Mpc, one need not consider the evolution of the event rate as a function of redshift for this estimate. Note that the numerical simulations in this work are performed assuming design sensitivities~\cite{dcc_public_ligo_sensitivity} for all four detectors.

\subsection{Improvement accounting for duty cycle} 

The above estimate is valid when all detectors take quality data simultaneously, which is rarely the case. However, it will be very difficult to identify the host galaxy of a merger if at least three detectors do not observe the event (or, rather, three detectors with ``similar'' sensitivity are observing simultaneously, even if all of them did not ``observe'' the event). For instance, Virgo did not detect GW170817, but was still useful for sky localization~\cite{gw170817_event}. If the duty cycle for each detector is $p \sim 70\%$ (for the ongoing O4b run, it is $60-75\%$~\cite{obsplan}), the probability that three of the four detectors will take quality data is $34\%$ (which would mean a factor $\sim 3$ increase in the required observation time to detect a specific number of events), while that for three out of four detectors is $65\%$. Thus, including LIGO-India in the network nearly doubles the time for which three detectors produce quality data simultaneously~\cite{Pankow_2020}.

This calculation can be extended to estimate the combined effect of the duty cycle and increase in the horizon distance, considering the probabilities of various possibilities (how many detectors operate simultaneously with what sensitivity) as a function of $p$ and $\alpha$. For any reasonable choice of these parameters, $\alpha = 25\%$ to $70\%$ and $p \approx 70\%$, the estimate shows that {\em LIGO-India will provide a factor $\sim 2.5$ times increase in the total number of (``triple-coincidence'') BNS merger detection localized for meaningful EM follow-up}. Inclusion of KAGRA in the estimation, with the horizon distance of $128$~Mpc~\cite{dcc_public_ligo_sensitivity} when LIGO-India starts to operate with A+ sensitivity, would change this increment factor due to LIGO-India from $\sim 2.5$ to $\sim 2$. It is also worth noting that if LIGO-India starts observing science data at half the target sensitivity, while the two other LIGO and Virgo detectors are operating at their next-level target sensitivity, and all detectors are operating at $70-80\%$ sensitivity, the improvement factor still stays close to $\sim 1.5 - 2$ due to the boost in the duty cycle. This is because the primary benefit of adding LIGO-India to the network arises from the increased duty cycle for triple-coincidence detections, rather than from a significant increase in SNR. Which is why, if LIGO-India starts operating at a lower sensitivity, the overall increase in the detection rate remains large.

The discussion above shows that LIGO-India can cut down the time required for a precise estimation of the Hubble constant by more than a factor of two without considering sky localization improvements (that is, without taking into account the geographical advantage of LIGO-India for sky localization). However, in practice, the time required will be much less. Here, we do not consider that detectors will gain significantly more sensitivity as time progresses. In this work, we are more interested in the relative improvement due to LIGO-India. Predicting future enhancements in the detector sensitivities is challenging and may have a minimal impact on the ratio of time required to detect several events with and without LIGO-India. Moreover, the period during which detectors are upgraded contributes to a delay in estimating the Hubble constant, effectively reducing the duty cycle. However, since planned upgrades are synchronized, the probability of multiple detectors operating simultaneously during an observing run remains unchanged.

\subsection{Improvement in event localization}
\label{subsec:skyloc}

\begin{figure}[ht!]
\hspace*{0.025\textwidth}
    \includegraphics[height=.3\textwidth]{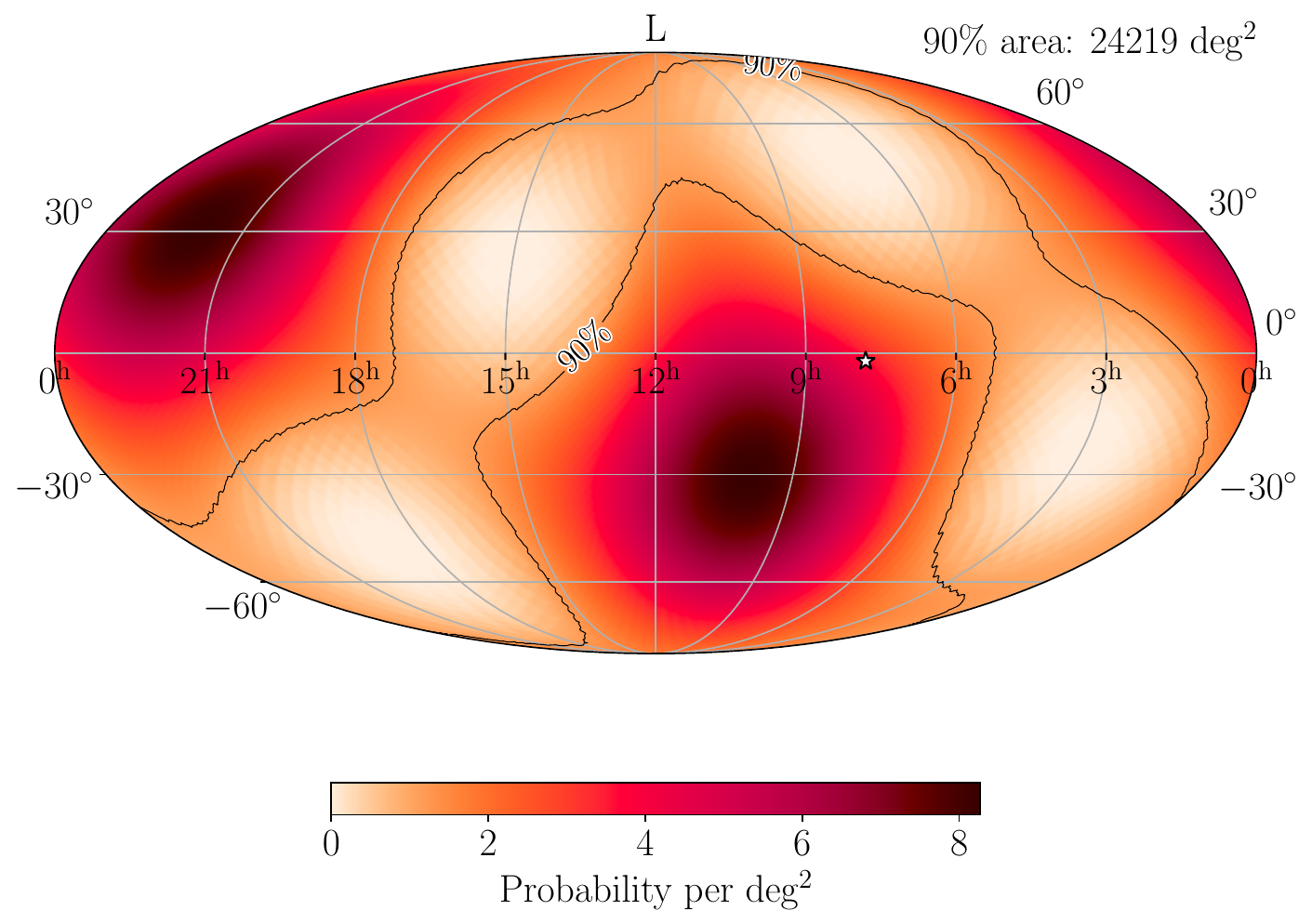} %
    \includegraphics[height=.3\textwidth]{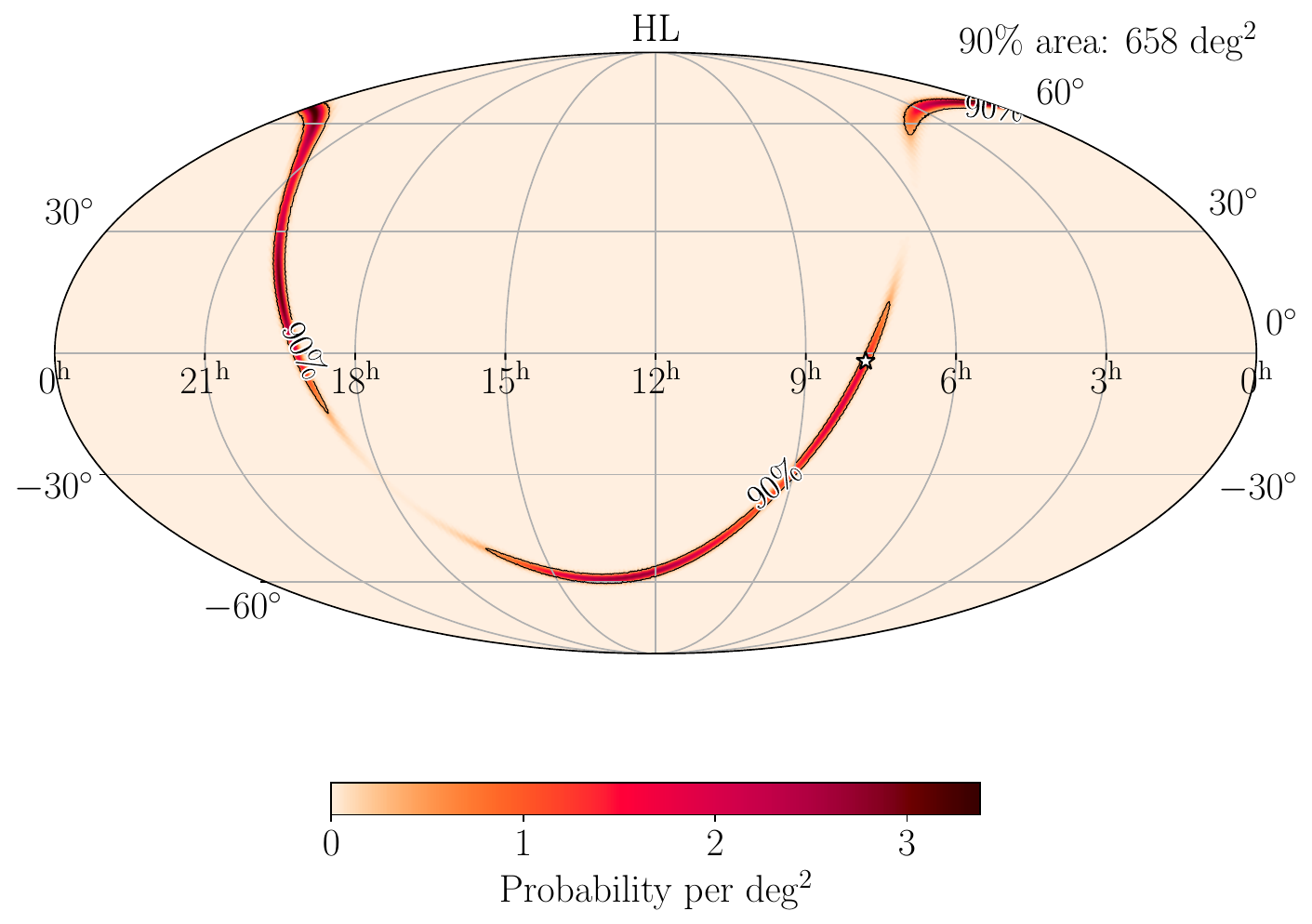}\\
    \hspace*{0.025\textwidth}
    \includegraphics[height=.3\textwidth]{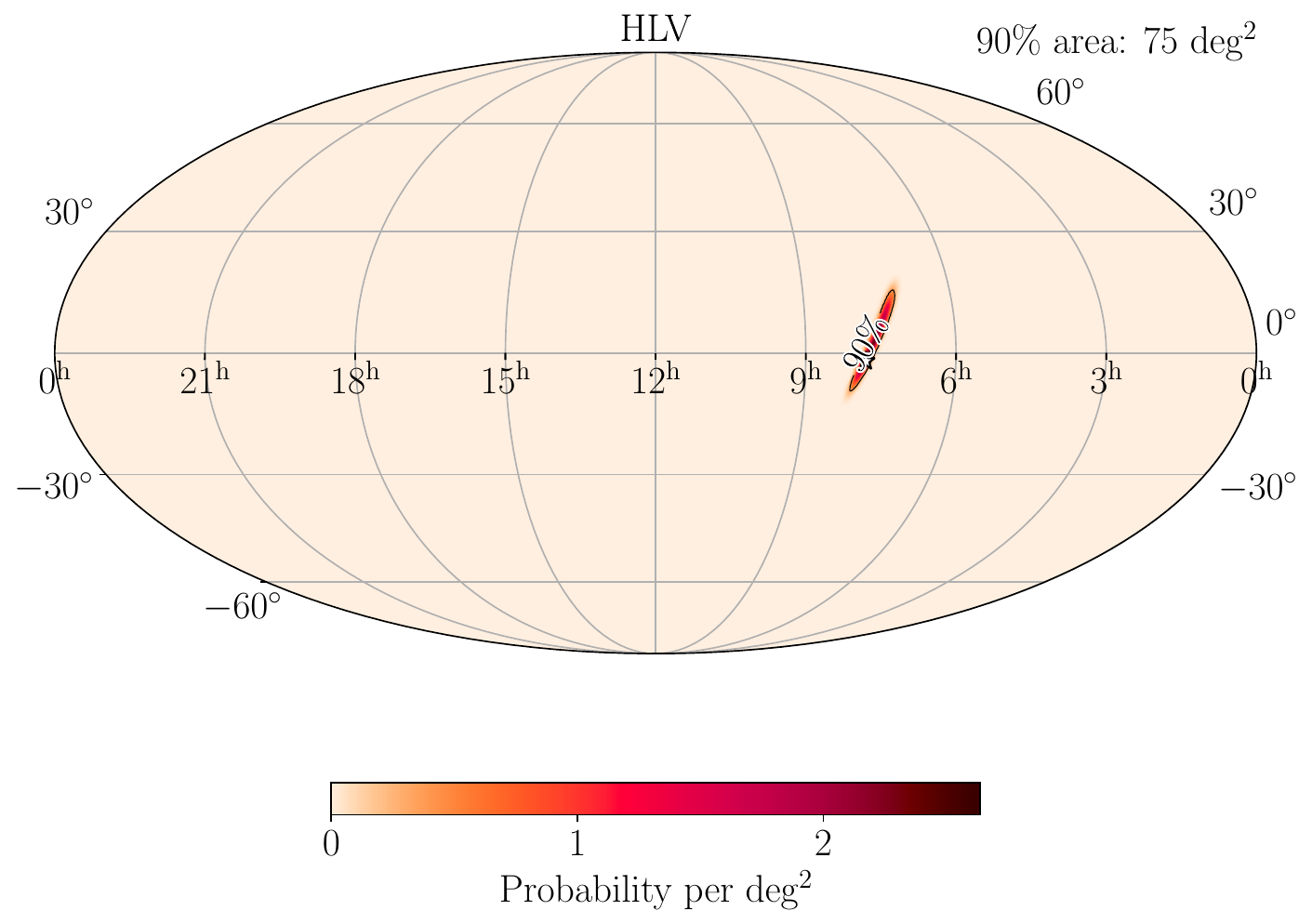} %
    \includegraphics[height=.3\textwidth]{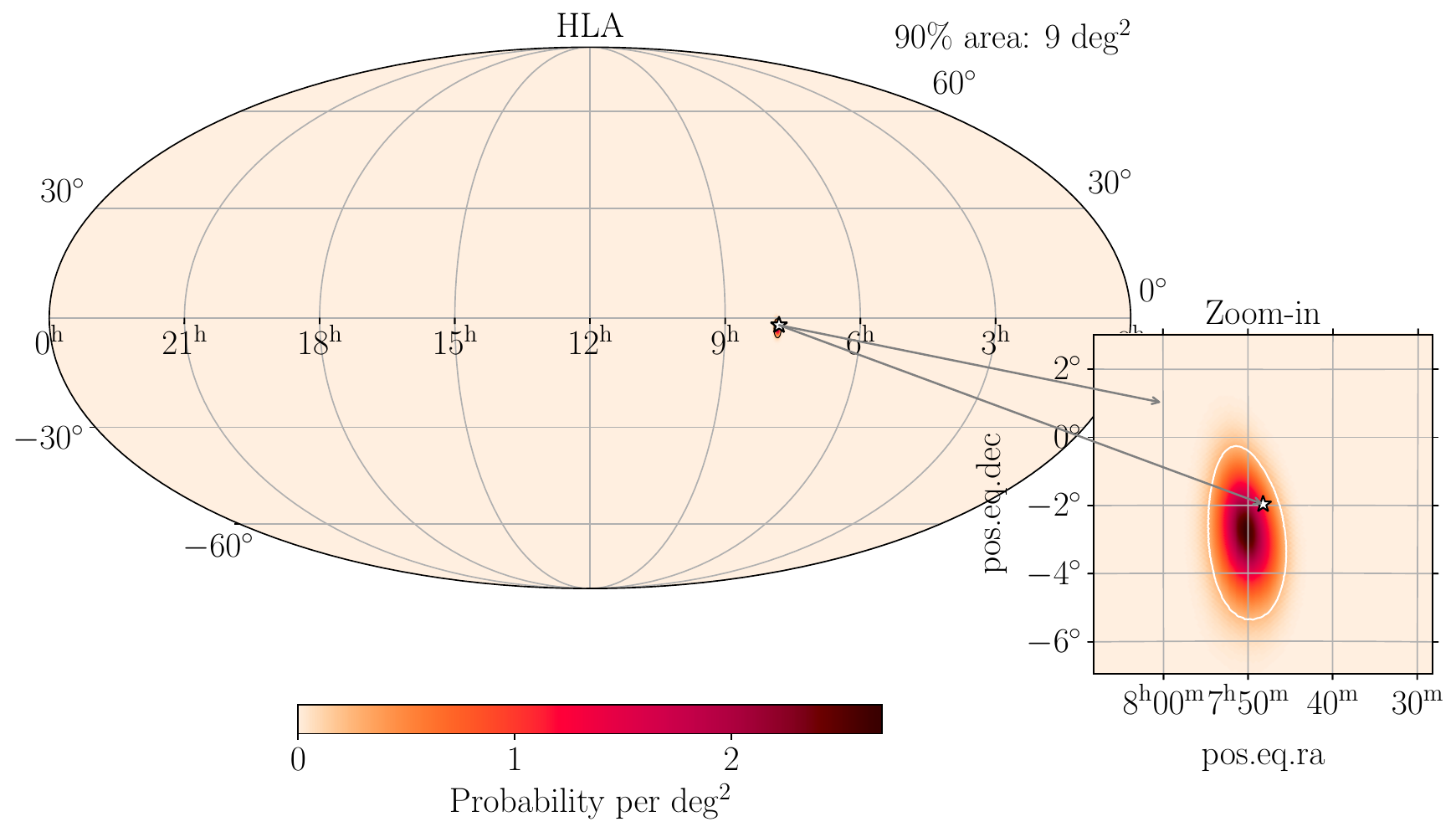}
    \caption{90\% credible regions obtained from \texttt{BAYESTAR}~\cite{singer_bayestar2016} for a simulated BNS source with component masses of 1.4 $\mathrm{M}_{\odot}$ each, spin magnitudes $s_{1z}$ and $s_{2z}$ of 0.03 each, located at a sampled distance of 361.23 Mpc and inclination angle ($\theta_{Jn}$) 2.69 radians from Earth. A star marker denotes the source's location in the sky. The localization is computed for a single Livingston (L) detector (top left), a network comprising Hanford (H) and Livingston (top right), Hanford and Livingston with Virgo (V) (bottom left), and Hanford, Livingston with LIGO-India (A) (bottom right). We assume that the two LIGO detectors and LIGO-India have A+ sensitivity, and Virgo has Advanced Virgo Plus (AdV+) sensitivity~\cite{dcc_public_ligo_sensitivity} for generating these sky localizations. Note that the addition of LIGO-India (even without Virgo) leads to a sky-localization area comparable to the FOV of LSST, implying that the telescope could spend all the available time on a single tile for follow-up, going much deeper in magnitude and significantly increasing the chances of observing distant or not so favorably aligned events.}
    \label{fig:skmaps}
\end{figure}

Since the addition of the LIGO-India detector to the HLV network doubles the number of ``baselines'' (pairs of detectors) and the new baselines are one an average longer by a factor of $\sim 1.5$, the localization area (which can be compared to the diffraction limit of the network), decreases by a factor of few~\cite{dcc_public_scitific_benefits_of_LI,ligoindiaFairhurst,Saleem_2022_ligoindia,Schutz:2011tw}. This is demonstrated numerically in Fig.~\ref{fig:skmaps}. Again, if LIGO-India starts operating at a lower but similar sensitivity compared to the other detectors in the network, it will still provide a major boost to event localization, as was shown in a recent work~\cite{PhysRevD.109.044051}. It is worth noting in this context that when GW170817 was detected, the BNS horizon distance of Virgo was $58$~Mpc while that for the LIGO Livingston and Hanford detectors were $218$~Mpc and $107$~Mpc respectively and the SNR of the event in Virgo did not even cross the detection threshold, yet, Virgo's presence with a ``similar'' sensitivity level reduced the sky localization error by more than a factor of $6$~\cite{gw170817_event}.

Enhanced localization precision increases the chances of observing EM counterparts during the EM follow-up campaign of a GW event. The sky area where an event is localized must be tiled and observed by an EM telescope. Typically, the size of these tiles is smaller than that of the telescope's field of view (FOV). If the product of the number of tiles and the exposure time per tile is not excessively large, the event can be followed up quickly before it fades or sets.

Considering the aforementioned points, we can see that LIGO-India is expected to increase the number of EM counterpart observations of BNS mergers by at least $\sim 50\%$, and quite likely by a factor of a few. Moreover, precise localization of the source can enable a telescope to observe each tile many times and for extended periods, thereby probing deeper and increasing the chances of observation of the EM counterpart by many folds. This important aspect has been quantitatively demonstrated in the next section (see  Fig.~\ref{fig:emfollowupresults}).

\section{Method}\label{sec:method}

\begin{figure*}[ht!]
    \centering
    \includegraphics[width=\textwidth]{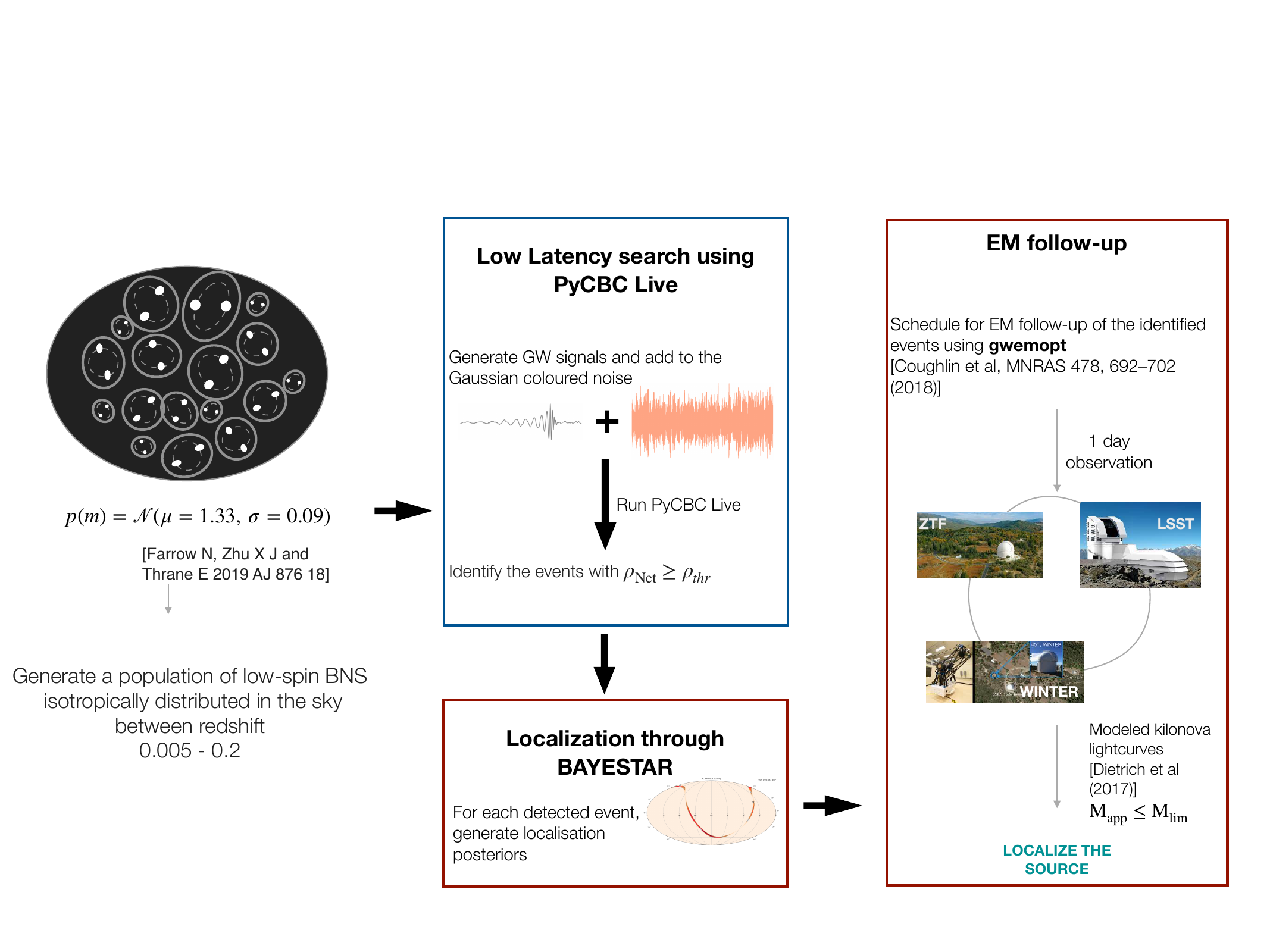}
    \caption{Flowchart illustrating the end-to-end simulation process used in our study. The first column shows the generation of a population of BNS sources, as described in Sec.~\ref{sec:simulation}. The middle column represents a realistic scenario where a low-latency search is performed on mock data containing simulated injections using \texttt{PyCBC Live}, followed by generating localization skymaps with \texttt{BAYESTAR} for sources that meet the detection criteria, as detailed in Sec.~\ref{sec:pycbclive}. The third column focuses on the kilonova follow-up of detected sources using the ZTF, LSST, and WINTER telescopes. A source is localized if its apparent magnitude ($\rm M_{app}$) is less than or equal to the telescope's limiting magnitude ($\rm M_{lim}$). Image sources: ZTF~\cite{ztf_image}, LSST~\cite{lsst_image}, and WINTER~\cite{winter_2020}.}
\label{fig:simulation_setup}
\end{figure*}

\subsection{Mock data analysis and simulation}\label{sec:simulation}

We simulate an observation scenario of the three LIGO detectors (Livingston, Hanford, and Aundha) and the Virgo detector spanning one year. Our primary focus is to assess the detectability of BNS mergers. Within this setup, we explore different possible operating conditions for the detectors. The first, which offers a more pragmatic view, assumes that each of the four interferometers operates with a duty cycle of 70\%. This number is chosen based on the third observing run of the two LIGO and Virgo detectors~\cite{gwtc3}. For comparison, we also examine scenarios where all detectors operate at full and half-duty cycles, 100\% and 50\%, respectively.

\begin{figure}[ht!]
    \centering
    \includegraphics[scale=0.35]{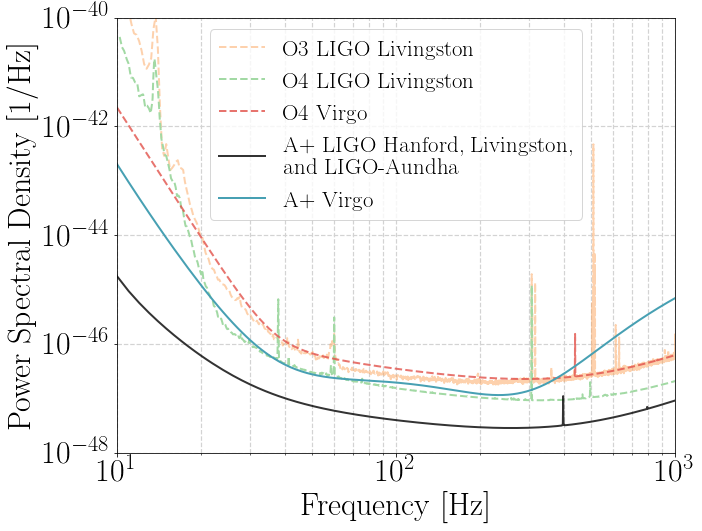}
    \caption{Projected Power spectral densities (PSDs) for A+ LIGO Hanford, LIGO Livingston, LIGO-India, and Virgo AdV+ are presented. For comparison, the observed PSDs from the O3 run and projected PSDs for the O4 run of LIGO Livingston and Virgo are also included. PSD data is taken from Ref.~\cite{dcc_public_ligo_sensitivity}.}
    \label{fig:psd_curves}
\end{figure}

We begin the simulation (see Fig.~\ref{fig:simulation_setup}) by first generating Gaussian noise realizations for each interferometer following their respective PSDs, as illustrated in Fig.~\ref{fig:psd_curves}. We use the publicly available LIGO A+ and Advanced Virgo Plus (AdV+) sensitivity curves~\cite{dcc_public_ligo_sensitivity}. Given our objective to identify GW signals from BNS mergers, we construct a realistic population of BNS sources and incorporate GW signals from their mergers into the noise strains.

\begin{figure}[ht!]
    \centering
    \includegraphics[scale=0.35]{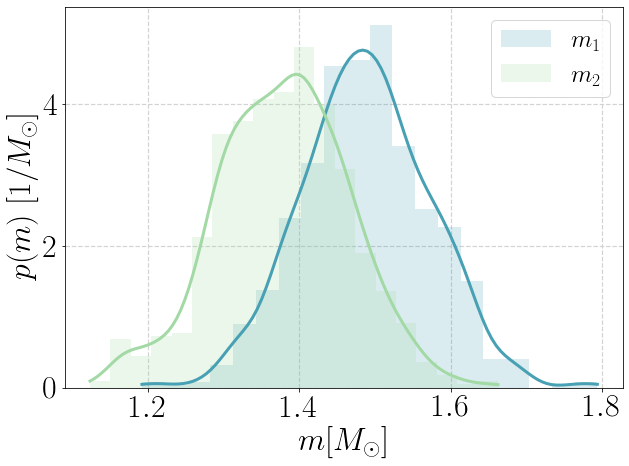}
    \caption{The probability density function of the primary ($m_1$) and secondary ($m_2$) source-frame masses of the simulated BNS signals. }\label{fig:mass_distribution}
\end{figure}

Following Ref.~\cite{Farrow_2019}, we simulate a population of BNS sources with neutron star mass following a well-fitted Gaussian distribution with a mean 1.33 $M_{\odot}$ and a width of 0.09 $M_{\odot}$\footnote{The chosen distribution describes neutron stars in the Milky Way, but is inconsistent with GW observations~\cite{KAGRA:2021duu} of BNSs which also include support for higher masses; however, since the allowed range of neutron star masses is relatively narrow, we do not expect the detectability of BNSs and other results to be significantly affected by this choice.}. Furthermore, we assume that the NSs in these binary systems exhibit slow rotation, resulting in a dimensionless spin component ranging from 0 to 0.05. The distribution of primary and secondary masses for the simulated sources is depicted in Fig.~\ref{fig:mass_distribution}. These injections were distributed isotropically in the sky and uniformly in comoving volume and source frame time, with a redshift range of 0.005 to 0.2. We keep the tilt angle $\theta_{JN}$, which is the angle between the total angular momentum and the observer's line of sight, uniform in $\sin\, (\theta_{JN})$. For simplicity, we exclude the influence of the orbital phase at coalescence and the polarization angle. We used \texttt{SpinTaylorT4}~\cite{spintaylor} model to generate the injections. As a result, we generate approximately 767 BNS injections with comoving distances of up to 500 Mpc. The discussions on relative improvement with the inclusion of LIGO-India in the forthcoming sections are based on these injections, but we scale all numbers to realistic absolute rates~\cite{KAGRA:2021duu} whenever required. Note that we do not include the effect of the peculiar velocity of the GW signal on the source, both in our mock data generation and while estimating parameters from the injections. Marginalizing over peculiar velocities could increase the uncertainty by a few per cent in recovering the Hubble parameter from standard sirens~\cite{Mukherjee:2019qmm}. 

Every generated signal was embedded into a 1024-second Gaussian noise segment, sampled at 8192 Hz, ensuring that the entire signal is captured within these segments. Due to the computationally intensive nature of generating such long-duration data, the segments were subdivided into 32-second intervals with appropriate padding at the boundaries. We uniformly applied this method across all four detectors, starting from a frequency of 15 Hz. 

\subsection{\label{sec:pycbclive}Low-latency search}

The search for GW signals emitted by compact binary mergers is commonly performed by employing the matched filtering technique~\cite{svd,Finn_1993,svd-sathya,svd-schutz,sathya-owen,findchirp}. In this technique, data from each detector in the network is correlated with a collection of accurately modeled waveforms known as templates. If the matched filter SNR crosses a certain threshold in all detectors, the event is identified as a detection\footnote{In reality, GW events are identified based on its false alarm rate (FAR) estimated using a noise background~\cite{usman}. However, thresholding on matched filter SNR is a good approximation to the behavior of searches~\cite{Essick:2023toz}.}. This technique is routinely used for low-latency searches and is implemented in \texttt{PyCBC Live}~\cite{alex_nitz_pycbclive}, a highly effective tool for efficient searches for GW sources in LIGO data. 

For our study, we use \texttt{PyCBC Live} to identify GW signals originating from BNS mergers within the simulated data. To search, we first construct a template bank that thoroughly explores the parameter space associated with BNSs. The template bank is generated by stochastically placing \texttt{SpinTaylorT4} waveforms, starting at 20 Hz~\cite{PhysRevD.80.104014}. We place these waveforms in a bank, ensuring that the loss in SNR remains below 3\%. Adhering to this criterion, we generate a bank of 36,191 non-precessing and quadrupolar templates, with their respective parameter ranges provided in Table~\ref{table:table1}.

\begin{table}[ht!]
\caption{Summary of the template parameter space in the bank. $m_1$ and $m_2$ represent the redshifted component masses of the NSs in a binary system. Their respective dimensionless effective spins are represented by $\chi_i$.} 
\vspace{0.2cm}
\centering
\begin{tabular}{
l@{\hspace{60pt}} c@{\hspace{40pt}} r@{\hspace{10pt}}}
\hline\hline
Parameter & Distribution & Range \\
\hline
$m_{1}, m_{2}$ & Uniform in detector-frame & 1.0--3.0 $M_\odot$ \\
$|\chi_1|, |\chi_2|$ & Uniform & 0--0.05 \\
\hline\hline
\end{tabular}
\label{table:table1}
\end{table}

We analyze the simulated data sampled at 4096 Hz using \texttt{PyCBC Live} and perform matched filtering with the templates generated at 20 Hz from the constructed bank. Currently, this pipeline is optimized to analyze data from a three-detector network in its current configuration. However, for our study, we modified the search pipeline to accommodate a four-detector configuration. In the analysis, we adopt a criterion for trigger identification: a signal must exhibit a matched-filter SNR $\rho$ greater than 4.5 in each detector. Subsequently, the list of triggers undergoes a coincidence test, wherein a time coincidence is observed across the detectors.

For detection, we estimate a quadrature sum ($\hat{\rho}_{c}$)\footnote{$\hat{\rho}_{c} = \sqrt{\sum_{i} \rho_{i}^2}$, where `i' runs for the number of detectors in the network.} of $\rho$ in each detector. We identify a true event in the three-detector scenario if $\hat{\rho}_{c} \ge 12$. We do not compute the false alarm rate in our study, as estimating the background and determining the $p$-value for detections in a four-detector network poses significant challenges and is beyond the scope of this work. To model a duty cycle of (say) $p\%$, we assume that there is a $p\%$ chance that a detector is observing at a given epoch. If a detector is not observing at the time of a given event, we remove that detector from our analysis. We also assume that the downtimes are not correlated between the detectors.

\subsection{Electromagnetic follow-up}\label{sec:em_followup}

To follow up on GW events that satisfy our detection criteria and detect associated kilonovae, we employ \texttt{gwemopt}~\cite{coughlin2018,coughlin2019} for planning and scheduling the EM follow-up campaign. This toolkit utilizes the spatial probability skymaps associated with GW events generated using a low-latency sky localization toolkit called \texttt{BAYESTAR}~\cite{singer_bayestar2016}. First, a skymap is subdivided into tiles that cover the FOV of the telescope. Then in each tile, the integrated spatial probability $T_{ij}$ is calculated as a function of the right ascension ($\alpha$) and the declination ($\delta$) as 

\begin{equation*}
    T_{ij} = \int^{\alpha_i + \Delta \alpha_i}_{\alpha_i} \int^{\delta_i + \Delta \delta_i}_{\delta_i} L_{\rm GW}(\alpha,\delta) \, d\Omega
\end{equation*}

for the event's sky location probability $L_{\rm GW}(\alpha,\delta)$. 

\begin{table}[ht!]
\centering
\caption{Telescope configurations of the ZTF, LSST, and WINTER used for the EM follow-up campaign using \texttt{gwemopt}. The choice of exposure times for each filter is consistent with Ref.~\cite{coughlin2018}.}
\begin{tabular}{l c c c c r}
\hline\hline
\addlinespace
Telescope & FOV & FOV shape  & Exposure time & Filter & Limiting \\ 
& (deg$^2$) & & (sec) & & magnitude \\
\addlinespace
\hline
\addlinespace
ZTF & 47 & Square & 30 & r & 21.10 \\ 
LSST & 9.6  & Circle & 30 & r & 24.25 \\ 
WINTER & 1   & Square & 450 & J & 21.00 \\ 
\addlinespace
\hline\hline
\end{tabular}
\label{table:telescopes}
\end{table}

For this study, we adopt the multi-order coverage tiling scheme~\cite{GRECO2022100547} to efficiently tile the skymaps in the EM follow-up simulation. We followed up a source for one day using the ZTF, LSST, and WINTER telescopes to detect their kilonova counterpart. These telescopes are chosen to observe both visible emissions and late-time infrared emissions from the identified GW events. Specifically, we utilized the r-band from ZTF and LSST, and the J-band from the WINTER telescope. To ensure effective source detection and localization, we employ specific exposure times for each telescope, allowing them to reach their standard magnitudes. The detailed exposure times for each telescope are provided in Table~\ref{table:telescopes}.

Accurate localization of a source within a specific tile requires the source's apparent magnitude to be brighter than the limiting magnitude of the telescope, as listed in Table~\ref{table:telescopes}. To determine the apparent magnitude of a given source, we simulate its kilonova light curve. This modeling process is elaborated upon in the next section.

\subsection{Lightcurve modeling}\label{sec:simulating_lightcurve}

Kilonovae emit a time-varying flux. The lightcurves of kilonovae early-time (minutes to hours) radiation have peaks in the ultraviolet wavelengths, shifting to redder wavelengths, and having a peak in the radio band at late times (months to years). Various models for the lightcurves have been proposed; they all differ in the physics prescriptions they employ and predict different lightcurves. Since we only have observations of a handful of kilonovae, model selection amongst these candidate models is impossible.




We use the prescription in Ref.~\cite{2014MNRAS.439..757G} to calculate the bolometric lightcurve, assuming a constant ejecta mass $M_{\rm ej} = 10^{-2} \, M_\odot$ and a constant ejecta velocity $v_{\rm ej} = 0.1 c $ consistent with the observation of the kilonova accompanying GW170817. Using the implementation of this model in the \texttt{gwemlightcurves}~\cite{gwemlightcurves_coughlin_2018,gwemlightcurves_coughlin_2019} package, we simulate the observed lightcurves in three different telescopes. The values of $M_{\rm ej}$ and $v_{\rm ej}$ are dependent on the equation of state (EOS) of NSs, which is as yet ill-constrained from observations. Instead of assuming constant $M_{\rm ej}$ and $v_{\rm ej}$, we could have assumed a fixed EOS for our calculations---a soft EOS would have led to more luminous (i.e., easier to detect) kilonovae, and vice versa for a stiff EOS~\cite{2017CQGra..34j5014D}.

\subsection{Source localization}\label{sec:source_localization}

For the obtained lightcurves, we calculate the apparent magnitude in each tile for a source. A source is considered detected if its apparent magnitude is lower than the limiting magnitude of a telescope. To determine the number of sources that are successfully localized by a telescope, we sum the probabilities associated with all the events recorded by the tiles using two distinct approaches. These two approaches were chosen to account for different criteria in source localization. 

First, we sum the probabilities of only the tiles with the highest probability of identifying sources. This approach can be mathematically represented as $\sum_{s} \max T_{ij, s}$. Second, we sum all the probabilities recorded by all tiles for all sources, which means $\sum_{s} T_{ij, s}$. The former method prioritizes tiles that are most likely to contain the source, thereby focusing on the most probable detections. However, given the inherent difficulty in accurately pinpointing the exact position of a source, the latter method is generally more reasonable, as it provides a cumulative likelihood of source localization over the entire observation field, offering a more comprehensive assessment.

\section{Results}\label{sec:results}

\begin{figure}[ht!]
    \centering
    \includegraphics[width=0.45\textwidth]{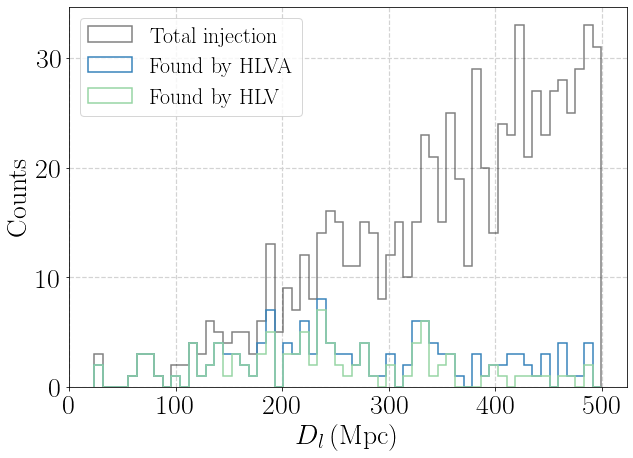}
    \includegraphics[width=0.48\textwidth]{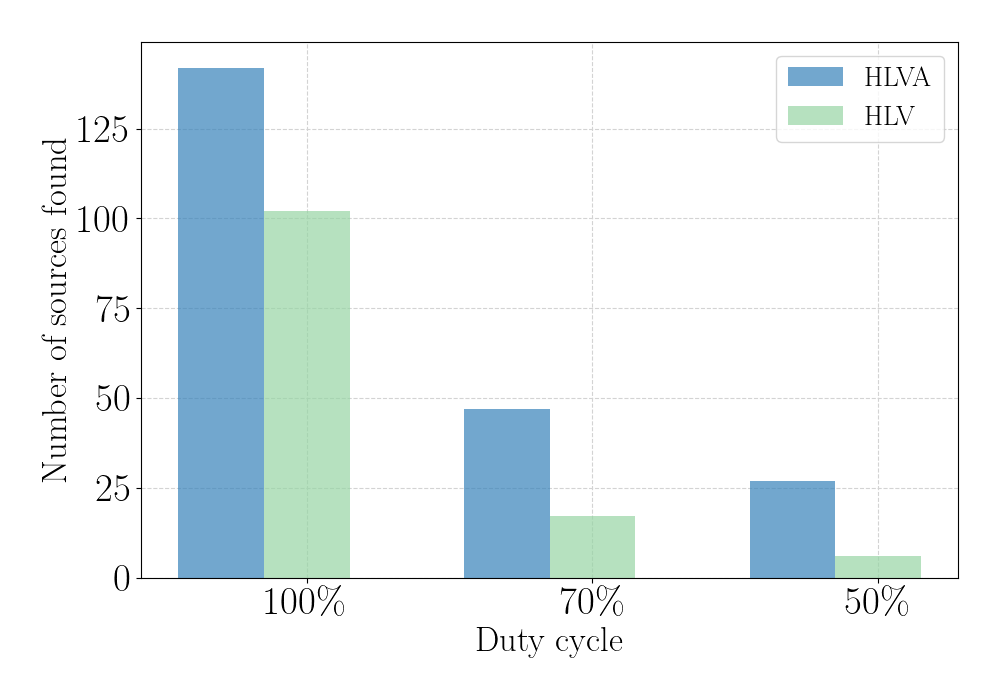}
    \caption{Plot illustrates the number of BNS sources identified using \texttt{PyCBC Live}. The left panel presents a histogram of the detected source's luminosity distance distribution ($D_L$). The blue and green curves represent the distributions for sources detected by the HLVA and HLV networks, respectively, assuming a 100\% duty cycle for all four detectors. For reference, the distribution of all injected sources is shown in gray. The right panel displays the total number of sources detected by the HLVA and HLV networks at 100\%, 70\%, and 50\% duty cycles.}
    \label{fig:pycbc_live_result_hlva_hlv}
\end{figure}

We conducted an extensive low-latency search for simulated BNS injections using \texttt{PyCBC Live}, as detailed in Sec.~\ref{sec:pycbclive}, for both HLVA and HLV network configurations. Of 767 injections, we identified a total of 142 sources with three-detector network SNR at least 12 when the HLVA network was fully operational at a 100\% duty cycle. In contrast, 102 sources were detected with the HLV network in operation. The number of sources detected varied with the duty cycle. When applying the duty cycle constraint, as described in Sec.~\ref{sec:pycbclive}, our results show that with the inclusion of LIGO-India in the network, the detection rate increases by a factor of approximately 1.4 at 100\% duty cycle, 2.8 at 70\%, and 4.5 at 50\% duty cycle. This trend is also evident in Fig.~\ref{fig:pycbc_live_result_hlva_hlv}. These estimates of the percentage improvement are consistent with the numbers derived in Sec.~\ref{sec:boe}. Note that we have used the projected A+ sensitivity for all detectors in this analysis. Since the estimates provided in Sec.~\ref{sec:boe} are reliable, they can offer insights into potential scenarios if one of the detectors fails to reach the projected sensitivity.


\begin{figure}[ht!]
    \centering
    \includegraphics[scale=.254]{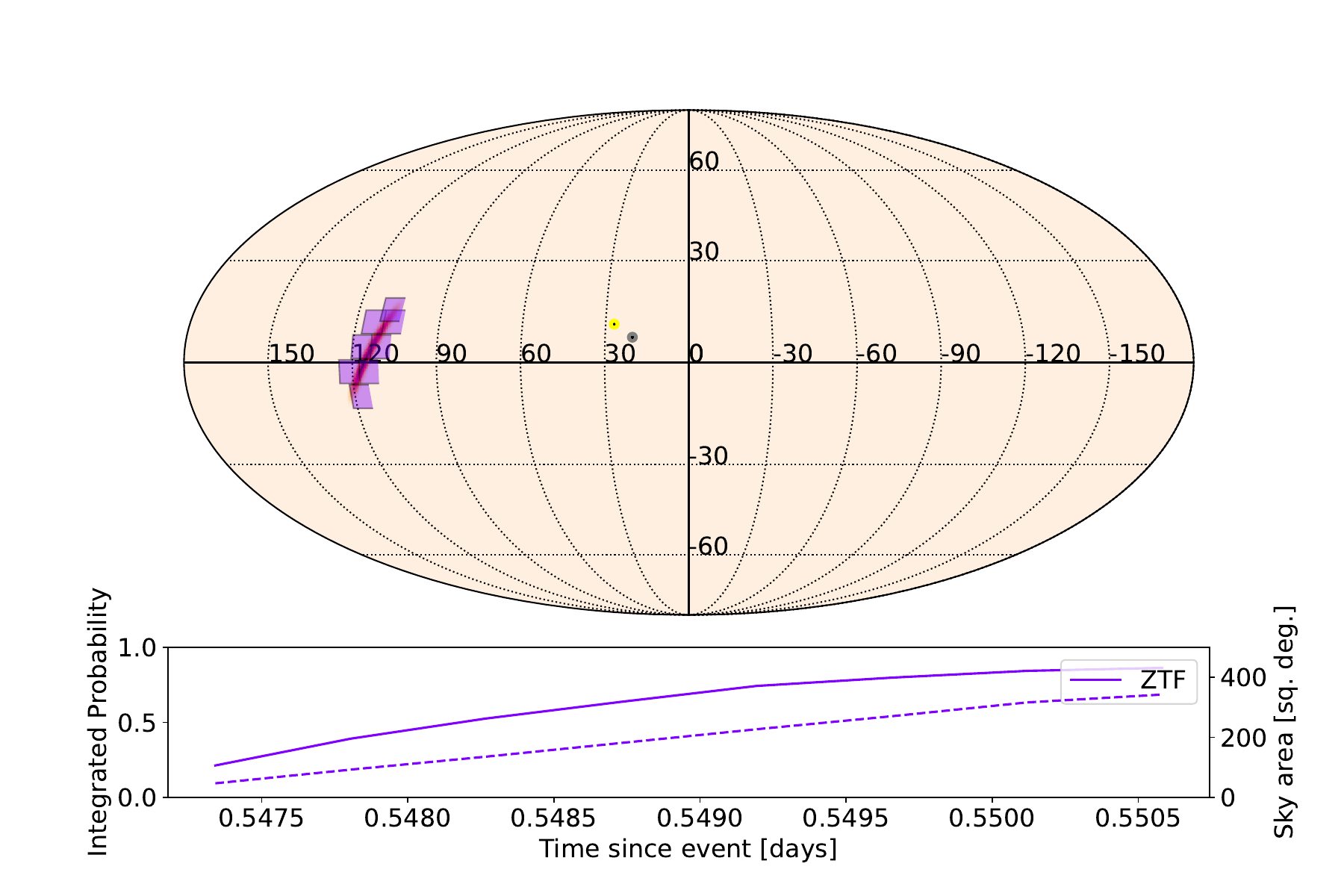}
    \includegraphics[scale=.254]{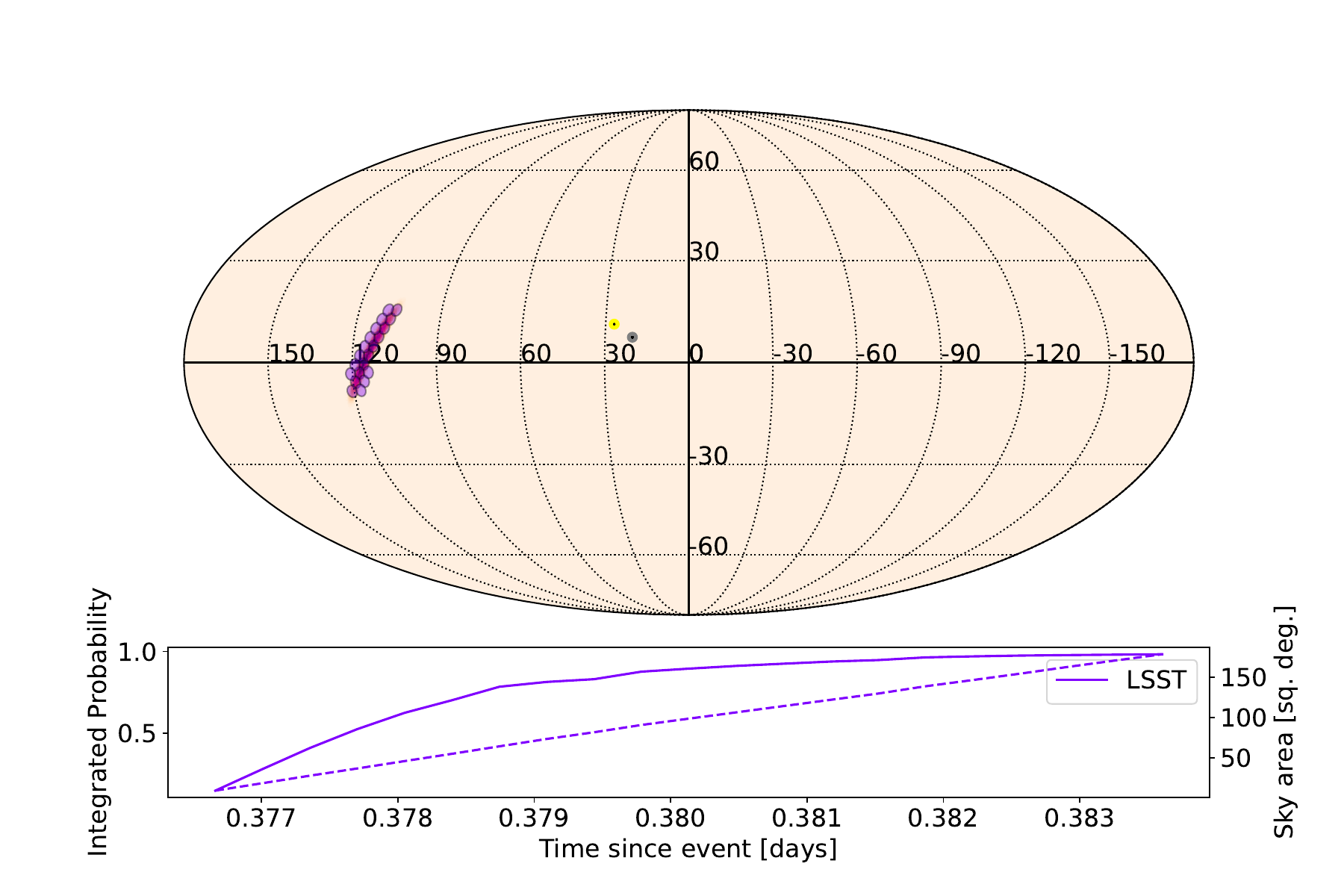}
    \includegraphics[scale=.254]{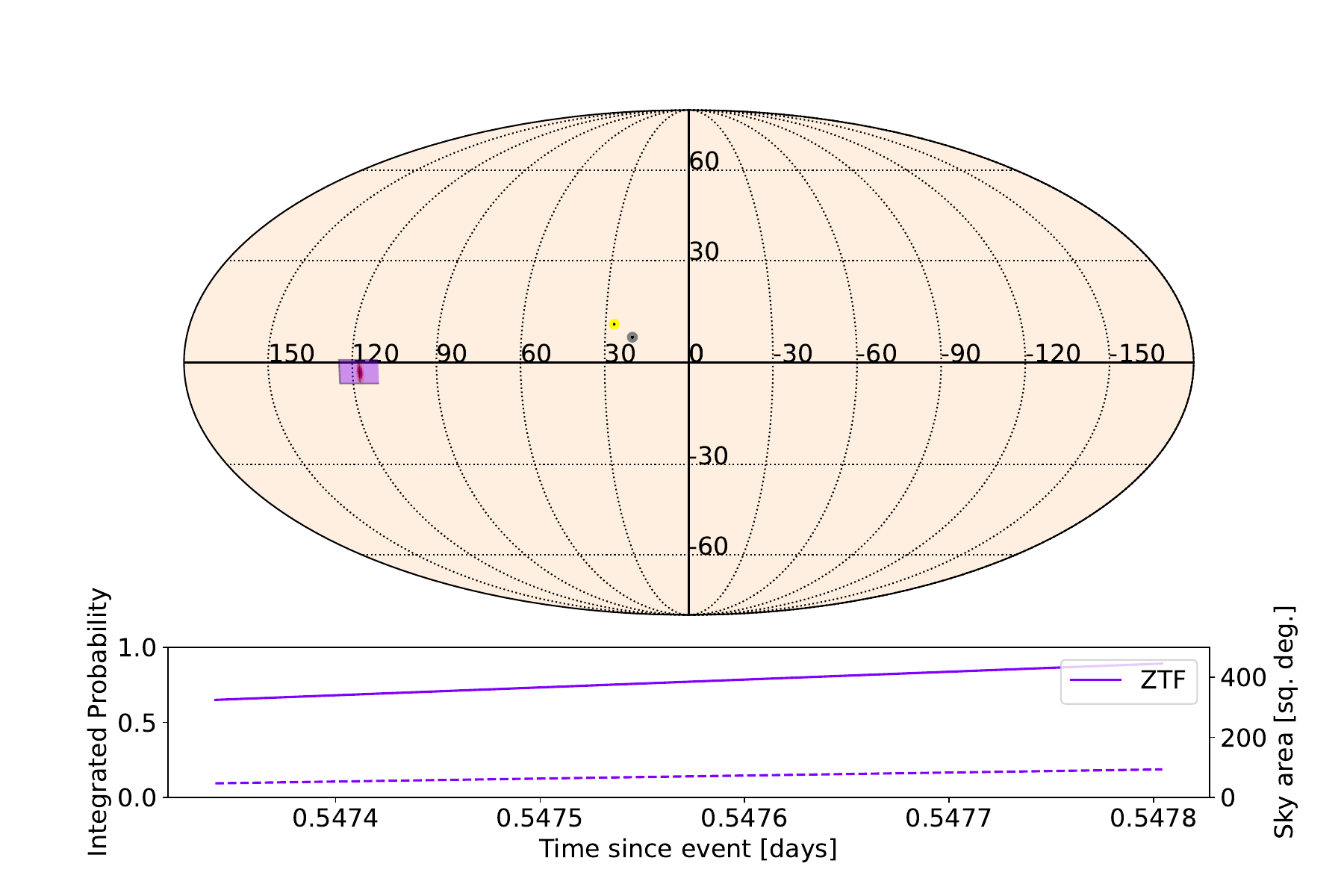}
    \includegraphics[scale=.254]{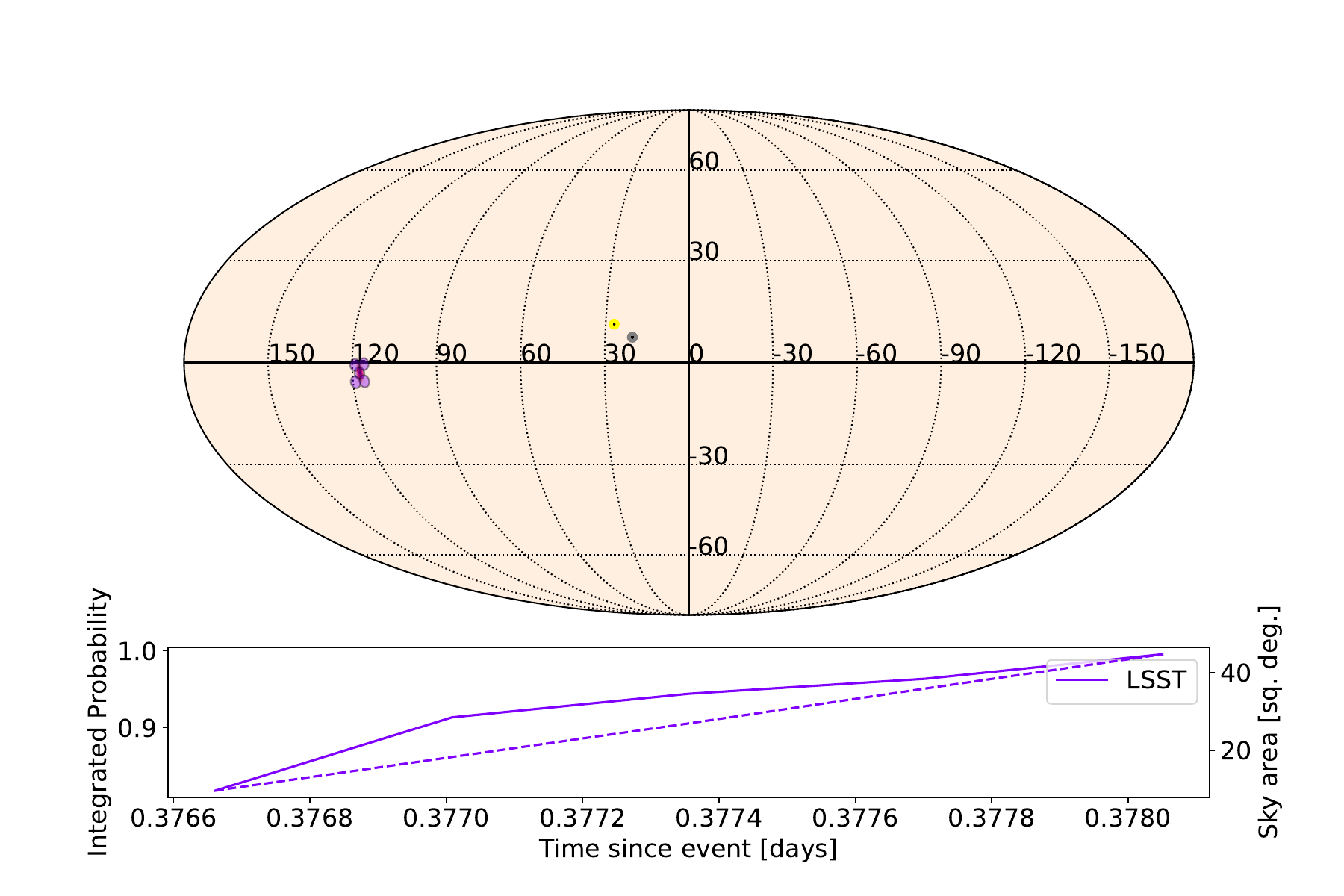}
    \caption{Optimized sky coverage for a simulated BNS source observed with ZTF (left panel) and LSST (right panel), using \texttt{gwemopt} for a single night of observations. Dashed lines indicate the total sky area covered, while solid lines represent the cumulative probability enclosed from the 2D skymap. The shaded area in magenta indicates the FOV-sized tile used to cover the sky area. The markers in yellow and grey are the positions of the sun and the moon, respectively. The ZTF tile is square, while the LSST tile is circular. For the HLV network (top panel), ZTF uses 8 tiles, and LSST uses 21 tiles. For the HLVA network (bottom panel), ZTF uses 2 tiles, and LSST uses 5 tiles. The plots are generated using \texttt{gwemopt}. This figure indicates the factor by which the EM telescopes can increase the exposure time for each tile, benefiting from precise sky-localization aided by LIGO-India to perform deeper follow-ups.}
    \label{fig:emfollowupresults}
\end{figure}

\begin{table}[ht!]
\centering
\caption{Total probability of detecting sources by telescopes, similar to Table~\ref{table:results_dutycycle_maxtileprob}. However, in this case, the probability of detecting a source within a tile is summed over all tiles for all sources. The results are presented for ZTF, LSST, and WINTER under different operating duty cycles of the HLVA and HLV (in parentheses) networks.}
\begin{tabular}{l c c r}
\hline\hline
\addlinespace
Duty cycle & ZTF & LSST & WINTER \\
\addlinespace
\hline
\addlinespace
100 \% & 9.55 (9.11)  & 60.68 (39.73) & 10.98 (9.44) \\
70 \% &  3.91 (0.68) & 20.93 (6.61) & 4.84 (0.61) \\
50 \% & 0.99 (0.94) & 10.05 (0.98)  & 1.94 (0.92) \\
\addlinespace
\hline\hline
\end{tabular}
\label{table:results_dutycycle_summedprob}
\end{table}

\begin{table}[ht!]
\centering
\caption{Total probability of detecting sources with apparent magnitudes below the limiting magnitudes of the three telescopes. The probabilities for all sources are summed only for tiles showing the peak probability of detecting a source. The results are presented for ZTF, LSST, and WINTER under different operating duty cycles of the HLVA and HLV (in parentheses) networks.}
\begin{tabular}{l c c r}
\hline\hline
\addlinespace
Duty cycle & ZTF & LSST & WINTER \\
\addlinespace
\hline
\addlinespace
100 \% & 6.17 (5.33) & 34.64 (14.87) & 4.14 (1.36) \\
70 \% & 2.11 (0.20) & 11.32 (1.98) & 1.66  (0.02) \\
50 \% & 0.77 (0.62) & 4.56 (0.46)  & 0.61 (0.15) \\
\addlinespace
\hline\hline
\end{tabular}
\label{table:results_dutycycle_maxtileprob}
\end{table}


\begin{figure}[ht!]
    \centering

    \includegraphics[width=0.475\textwidth]{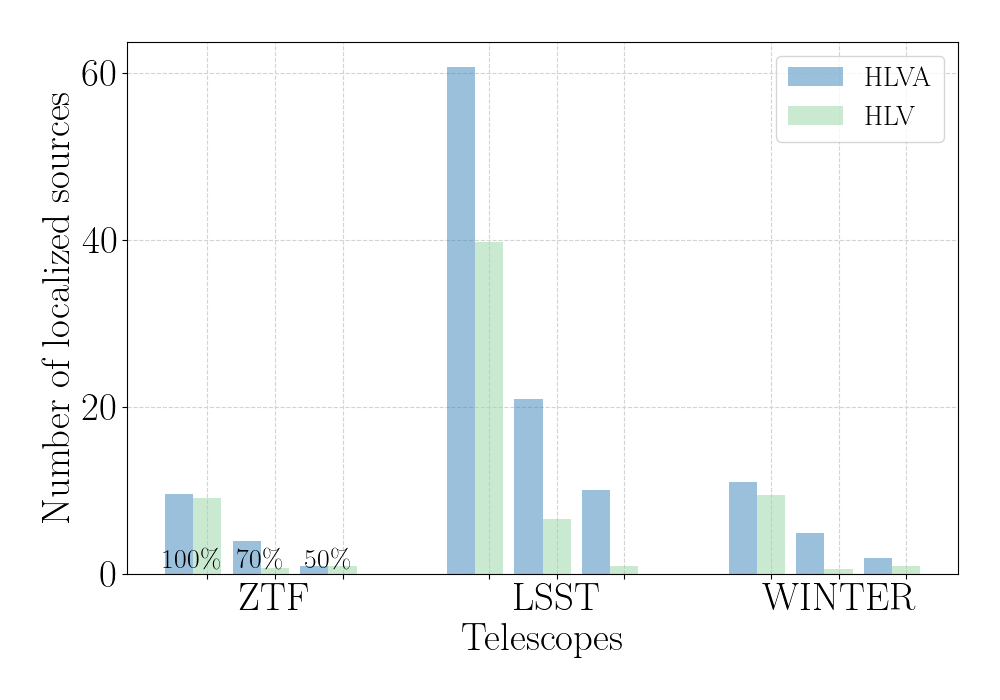}
    \includegraphics[width=0.475\textwidth]{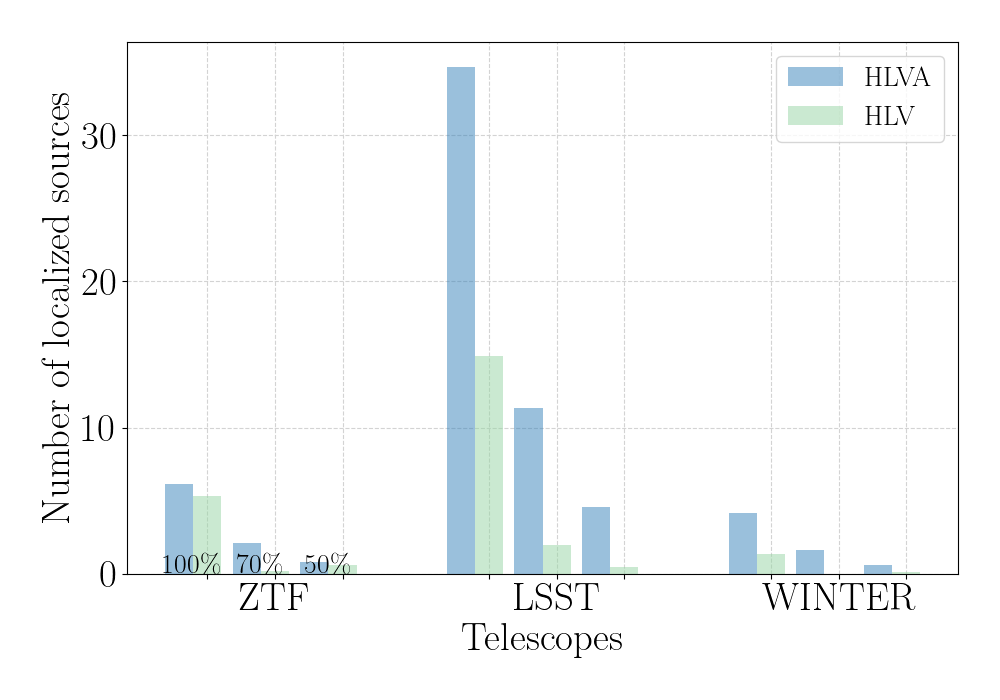}
    \caption{Comparison of source localization by ZTF, LSST, and WINTER telescopes with (in blue) and without (in green) the inclusion of LIGO-India (A) in the three-detector network across effective duty cycles of 100\%, 70\%, and 50\%. The first panel depicts the total probability of detecting sources across all tiles during a telescope's observation time. The second panel illustrates the total probability of source detection when observing tiles with the highest probability of locating the source within the telescope's FOV.}
    \label{fig:gwemopt_results1}
\end{figure}


When we followed up on these detected sources to identify their kilonova counterparts, we found that the number of localized sources was higher in the HLVA network compared to the HLV network, regardless of the localization approach used, as discussed in Sec.~\ref{sec:source_localization}. This result is evident in the two panels of Fig.~\ref{fig:gwemopt_results1} and Tables~\ref{table:results_dutycycle_maxtileprob} and~\ref{table:results_dutycycle_summedprob}.

The overall probability of detecting a source depends on how the source’s apparent magnitude diminishes over the specific number of tiles, making precise localization a tedious task. When focusing only on the tile with the highest probability of detecting a source, the cumulative probabilities of all sources show a significant improvement with the inclusion of LIGO-India compared to its exclusion from the network, as evidenced by the first panel of Fig.~\ref{fig:gwemopt_results1}. In this scenario, the inclusion of LIGO-India results in approximately twice the number of sources with better localizations compared to scenarios without LIGO-India, as shown in Table~\ref{table:results_dutycycle_maxtileprob}. This improvement becomes even more pronounced when LIGO-India operates alongside the other two LIGO detectors and Virgo at 70\% and 50\% duty cycles, yielding an overall improvement by a factor of three to five.

When all tiles are considered until the source’s apparent magnitude falls below the limiting magnitude, the total number of localized sources improves significantly compared to the scenario in which only one tile is considered, as shown in the second panel of Fig.~\ref{fig:gwemopt_results1} and Table~\ref{table:results_dutycycle_summedprob}.

While our primary work focus is on estimating the relative improvement in the EM follow-up rate with and without LIGO-India, it is possible to scale our results to obtain approximate absolute numbers according to the current expectations for the BNS merger rates. If one assumes that the merger rate is in the range $10-1000~\text{Gpc}^{-3}\text{yr}^{-1}$, the expected mergers within a volume of radius 500 Mpc (that is, $\sim 0.524~\text{Gpc}^3$) would be roughly $5-500$ per year. In our simulations, we injected $767$ BNS sources, which, when scaled to the above rate range, correspond to a period of $\sim 1.5 - 150$~years. If we consider the LSST scenario with $70$\% duty cycle (Table~\ref{table:results_dutycycle_summedprob}), the expected number of detections out of 767 injections with (without) LIGO-India is $20.93~(6.61)$. This implies that to detect $\sim 50$ BNS mergers, the expected number of years in the most optimistic cases will be $\sim 3.6~(11.3)$~years, and up to two orders of magnitude longer in the most pessimistic case. If we suppose the merger rate was say $100\, \text{Gpc}^{-3}\text{yr}^{-1}$, the corresponding timescale would be $\sim 36$ years with LIGO-India and $\sim 113$ years without it. If only single tile was observed, as depicted in Table~\ref{table:results_dutycycle_maxtileprob}, the timescales would increase further to $63$ and $379$ years, respectively. 

However, we emphasize that these numbers are intended only for illustrative purposes. The detectors are going to be upgraded in the coming years and the detection rates are expected to improve by a factor of few to one order of magnitude. An alternative way to interpret these results is by directly relating the EM follow-up rate to the underlying merger rate. If $N_\text{EM}$ detections are obtained from $N_{\rm inj}$ injections, then the astrophysical rate $R$ corresponding to these EM detections is given by 

\begin{equation}
    R = \frac{N_\text{inj}/N_\text{EM}}{\langle VT \rangle}\,, 
\end{equation}

where, ${\langle VT \rangle}$ is the volume-time sensitivity of the detector network. In the specific case of LSST with a 70\% duty cycle, the EM follow-up rate of 1 per year with HLV, corresponds to an actual astrophysical merger rate of $\sim 138~\text{Gpc}^{-3}\text{yr}^{-1}$.

The telescopes considered in this study have relatively short exposure times. This may significantly reduce the probability of detection of the EM counterparts, as discussed in the following sections. Consequently, the sky-localization improvement in the HLVA network may not significantly impact their observations. With short exposures, these telescopes can quickly cover a large sky area, as illustrated in the left panel of Fig.~\ref{fig:emfollowupresults}. We examined the improvement by focusing on the tile with the peak probability as a toy case to understand how sky localization could benefit telescopes observing deeper by increasing the exposure time or taking multiple exposures of the same tile. In this scenario, a telescope would observe only one tile with whatever observation time it can allocate. The results, shown in Table~\ref{table:results_dutycycle_maxtileprob}, indicate that the improvement is approximately a factor of 2.5 for LSST, even with a 100\% duty cycle of a four-detector network. This improvement becomes even more significant, around a factor of 5, when the duty cycle is 70\%, and increases further for lower duty cycles.

\begin{table}[ht!]
\centering
\caption{Fractional uncertainty in the Hubble constant ($\sigma_{H_0}$) estimated from BNS sources detected by LSST under different detector duty cycles. Values in parentheses correspond to the HLV detector network, while the others correspond to the HLVA network. The improvement factor quantifies the relative reduction in uncertainty when including the LIGO-India in the network.}
\vspace{0.5pt}
\begin{tabular}{lcc}
\hline\hline
\addlinespace
Duty Cycle  & $\sigma_{H_0} / H_0$ for HLVA (HLV) & Improvement Factor \\
\addlinespace
\hline
\addlinespace
100\%  & 0.0150 (0.0248) & 1.65 \\
70\%    & 0.0340 (0.0678) & 1.99 \\
50\%    & 0.0565 (0.1591) & 2.82 \\
\addlinespace
\hline\hline
\end{tabular}
\label{table:h0_em_telescopes}
\end{table}

In addition to detection statistics, we quantified the impact of LIGO-India on Hubble constant measurements using \texttt{BAYESTAR} and complementary back-of-the-envelope scaling for the followed-up events. In particular, we calculated the fractional uncertainty in distance measurement using \texttt{BAYESTAR} and propagated it to obtain the corresponding errors in the $H_0$ value. Table~\ref{table:h0_em_telescopes} summarizes the fractional uncertainty in the Hubble constant for different detector duty cycles using LSST follow-up observations. For a 100\% duty cycle, LSST achieves a low uncertainty of 0.0150 with the HLVA network, which increases to 0.0340 and 0.0565 for 70\% and 50\% duty cycles, respectively. Including LIGO-India consistently improves the measurement precision, with improvement factors ranging from 1.65 at full duty cycle to 2.82 at 50\%. 


\section{Discussion and Conclusion}\label{sec:conclusion}

GW observation and localization of tens of BNS merger events and EM follow-up may resolve the Hubble tension~\cite{chenmaya}. Although in principle this resolution could be achieved with the two LIGO and Virgo detectors by observing GWs from BNS mergers and their EM signals, it would take decades. In this work, we demonstrate that the combined advantages provided by LIGO-India can dramatically reduce this time, underscoring the enormous importance of LIGO-India for addressing critical questions in present astronomy.

However, certain assumptions underlie the conclusions of Ref.~\cite{chenmaya} and potentially introduce small biases in our understanding. Firstly, the merger rate of BNS with EM counterparts was assumed to be $1,540^{+3,200}_{-1,220}$ Gpc$^{-3}$ yr$^{-1}$, in line with findings of Ref.~\cite{gw170817_event}. Additionally, the study assumed that most BNS events would be associated with kilonovae, ensuring the identification of the host galaxy, which is crucial for $H_0$ measurement. Thus, the assumption in Ref.~\cite{chenmaya} about the number of detections required to achieve 2\% might lead to an overestimation. This possible overestimation comes from two key factors. First, current GW observations have not identified BNS sources other than GW170817 with an EM counterpart. This lack of observed EM counterparts suggests that the expected number of detectable events with EM counterparts may be lower than initially estimated. Second, observing a fast-fading kilonova in a limited observing time may not be possible in following up a source with an EM telescope. This limitation implies that a more extensive and deeper survey, possibly involving more sensitive telescopes or longer observation times, would be necessary to effectively increase the number of BNS detections helpful in refining the measurement of $H_0$. Considering these constraints and the current observational scenario of BNS events with counterparts, we assume the detection rate of BNS events with detectable counterparts to be one per year. 


The results presented in this paper indicate a significant increase in the detection capabilities of BNS mergers with their associated EM kilonova counterparts with the operation of LIGO-India. Specifically, we anticipate an increase in EM follow-up detection by approximately 2 to 7 times, with a realistic factor being around 4 to 5, as shown in Sec.~\ref{sec:results}. If one of the other detectors fails to achieve the projected sensitivity, this factor will increase, though the time required to achieve the target $H_0$ precision goal will also increase. Thus, LIGO-India should reduce the time required for a 2\% measurement of $H_0$ to about 10 years. This period will be further reduced with the increase in detector sensitivities as the LIGO and Virgo operate at A$^{\sharp}$~\cite{ahash_report} and Virgo nEXT~\cite{virgo_next} sensitivities, respectively. In particular, a 30\% improvement in sensitivity (horizon distance) implies a further reduction in this required period by a factor of $\sim 2$.

In this work, we have considered small exposure times (e.g., 30 seconds for LSST), but this time could be extended significantly. Especially when the sky localization area becomes comparable to the FOV of the telescopes after LIGO-India begins operating, telescopes should be able to spend the entire available observation time on very few tiles in the sky (see Fig.~\ref{fig:emfollowupresults}). This will lead to a significant increase in the observation of distant and not-so-favorably oriented sources. Considering that the horizon distance for EM follow-up increases as the square root of observation time and the volume coverage per unit solid angle increases as the cube of the horizon distance, a factor of a few increase in the observation time per tile due to precise source localization by LIGO-India can increase the probability of detection of EM counterparts by approximately one order of magnitude ($\sim 5^{3/2} \approx 11$). (Note that we have not included this factor in our simulations. Therefore, in our simulations, the total EM follow-up time is greater for poorly localized sources, proportional to the number of observation tiles.) This increase would be in addition to the boost in the GW detection probability of triple-coincidence localized BNS merger events, leading to an overall increase by a factor of $\sim 20$. This factor comes from the combination of an increase in the detection rate accounting for the duty cycle and the EM follow-up volume increment through deeper observations, which captures the effect of a sky-localization improvement. Therefore, it is reasonable to expect that the network of GW detectors, along with EM telescopes, will be able to resolve the Hubble tension within a few years of LIGO-India’s operation.

The primary uncertainty in our estimate comes from the observable rate of EM counterparts, here kilonovae, from BNS mergers. EM counterpart from a BNS event has been observed only once, and that too about seven years ago. Nevertheless, if we assume that the probability of observing such events is not negligible (although presently small), given that more BNS mergers were observed by the GW detectors after that (though no EM counterpart was detected) and that the number of detection of mergers is increasing rapidly (doubling every $\sim 1$ year~\cite{detection_plot_o3}), anticipating about $\sim 1$ per year of EM counterpart observation from BNS mergers in the A+ era seems reasonable. If this rate is much lower, we may have to wait for the third-generation detectors like the Einstein Telescope~\cite{Hild_2010,Punturo_2010,ET2017,Maggiore_2020,Pace_ET_2022}, and Cosmic Explorer~\cite{cosmic_explorer_Abbott_2017,Reitze:2019iox,cosmic_explorer_white_paper}, along with LIGO-India for precise enough sky-localization, to operate for a few years (perhaps in the 2040s) for resolving the Hubble tension through this window of astronomy~\cite{Branchesi_2023,evans2023cosmicexplorersubmissionnsf,gupta2024characterizinggravitationalwavedetector,Chen_Cosmography_2024,Pandey_2025}. On the other hand, if the rates are not so pessimistic and if LIGO-India begins operating by 2030 as planned, this breakthrough can happen by $\sim 2035$.

All files required to reproduce the simulation results and plots are available in the GitHub repository\footnote{\href{https://github.com/Kanchan-05/LIGO-India-and-hubble-trouble}{https://github.com/Kanchan-05/LIGO-India-and-hubble-trouble}}.

\section*{Acknowledgment}

The authors thank Stephen Fairhurst for a careful reading of the draft and helpful comments. The authors express their sincere gratitude for the computational resources provided by the IUCAA LDG cluster Sarathi and the support received from the LIGO Laboratory and National Science Foundation Grants. Additionally, the authors would like to thank Viraj Karambelkar for his invaluable insights during the initial phase of this work and Varun Bhalerao for his thoughtful discussions and constructive feedback, which helped refine this study. K. S. acknowledges the Inter-University Centre of Astronomy and Astrophysics (IUCAA), India, for support during the initial phase and the National Science Foundation Awards (PHY-2309240) for the final phase of this work. AV acknowledges support from the Natural Sciences and Engineering Research Council of
Canada (NSERC) (funding reference number 568580). S. M. acknowledges the Department of Science and Technology (DST), Ministry of Science and Technology, India, for the support provided under the esteemed Swarna Jayanti Fellowships scheme. This material is based on work supported by the NSF's LIGO Laboratory, a major facility fully funded by the National Science Foundation. This work utilizes open source packages \texttt{PyCBC Live}~\cite{pycbclive_zenodo}, \texttt{BAYESTAR}~\cite{singer_bayestar2016}, and \texttt{gwemopt}~\cite{gwemlightcurves_coughlin_2018,gwemlightcurves_coughlin_2019}. 

This manuscript has been assigned LIGO DCC Document No. P2400348.

\section*{References}
\bibliographystyle{iopart-num}
\bibliography{references}
\end{document}